\documentclass{aa}

\usepackage{amsmath}
\usepackage[latin1]{inputenc}
\usepackage[T1]{fontenc}
\usepackage{txfonts}

\usepackage{natbib}
\bibpunct{(}{)}{;}{a}{}{,} 

\usepackage[french,english]{babel}

\usepackage[breaklinks]{hyperref}
\usepackage{graphicx}

\usepackage{xcolor}

\providecommand{\unit}[1]{\ensuremath{\:\mathrm{#1}}}

\def\eg{{e.g.}\ }
\def\ie{{i.e.}\ }

\newcommand*{\ud}{\ensuremath\,\mathrm{d}}

\newcommand*{\Bo}{\ensuremath{\vec{B}_\parallel}}

\newcommand*{\Exp}[1]{\ensuremath{\mathrm{e}^{#1}}}
 
\newcommand*{\kperp}{\ensuremath{k_{\perp}}}

\newcommand*{\tnl}{\ensuremath{\tau_\textrm{NL}}}

\newcommand*{\tnu}{\ensuremath{\tau_\nu}}

\newcommand*{\shellatm}{\textsc{Shell-Atm}}
\newcommand*{\hydrad}{\textsc{HydRad}}

\begin{document}
\selectlanguage{english}

\title{Profiles of heating in turbulent coronal magnetic loops}

\author{E. Buchlin\inst{1} \and P.~J. Cargill\inst{1} \and S.~J.
  Bradshaw\inst{1} \and M. Velli\inst{2,3}}

\offprints{E. Buchlin, \protect\url{e.buchlin@imperial.ac.uk}}

\institute{Space and Atmospheric Physics Department, The Blackett
  Laboratory, Imperial College, London SW7 2BW, UK
  \and
  Dipartimento di Astronomia e Scienza dello Spazio,
  Università di Firenze, Largo E. Fermi 2, 50125 Firenze, Italy
  \and
  Jet Propulsion Laboratory, California Institute of
  Technology, 4800 Oak Grove Drive, Pasadena, CA 91109, USA
}

\date{Received\,:  / Revised date\,:}

\abstract{
  The location of coronal heating in magnetic loops has been the subject of
  a long-lasting controversy: does it occur mostly at the loop footpoints,
  at the top, is it random, or is the average profile uniform?
}{
  We try to address this question in model loops with MHD turbulence and a
  profile of density and/or magnetic field along the loop.
}{
  We use the \shellatm\ MHD turbulent heating model described in Buchlin \&
  Velli (2006), with a static mass density stratification obtained by the
  \hydrad\ model (Bradshaw \& Mason 2003).  This assumes the
  absence of any flow or heat conduction subsequent to the dynamic
  heating.
}{
  The average profile of heating is quasi-uniform, unless there is an
  expansion of the flux tube (non-uniform axial magnetic field) or the
  variation of the kinetic and magnetic diffusion coefficients with
  temperature is taken into account: in the first case the heating is
  enhanced at footpoints, whereas in the second case it is enhanced where
  the dominant diffusion coefficient is enhanced.
}{
  These simulations shed light on the consequences on heating profiles of
  the complex interactions between physical effects involved in a
  non-uniform turbulent coronal loop.
}


\maketitle

\section{Introduction}
\label{sec:intro}

It is now widely accepted that magnetic loops are the basic building block
of the closed solar corona in either active regions or the quiet Sun, and
the mechanism responsible for their heating constitutes a major unsolved
problem. The energy requirements in the corona are quite well known, with
approximately $300\unit{W m^{-2}}$ needed in the quiet Sun and $10^4\unit{W
  m^{-2}}$ in the active regions respectively \citep{withb77}.  It is also
believed that photospheric motions can provide this power, with this
injected energy being transported into the corona by waves or by the slower
movements of the magnetic field lines.  The difficulty lies in identifying a
credible dissipation mechanism, and it is clear that small scales ($<
1\unit{km}$), such as are likely to be produced by turbulence, are needed.
Following the work of \citet{hey92} and \citet{gom92} who showed that a
self-consistent model of turbulence could account for coronal heating, many
numerical simulations have been performed in order to study this scenario,
\eg by 2D direct numerical simulations of MHD \citep[e.g.][]{dmi98, geo98},
by cellular automata \citep[e.g.][]{luh91, vla95, buc03} or by shell-models
\citep[e.g.][]{nig04, buchlin07pre}.

An important aspect of such studies which is the subject of this paper
concerns the spatial distribution of the heating as a function of distance
\emph{along} the loop. A determination of average energy dissipation profile
as a function of the position along a coronal loop would put a strong
constraint on the physical mechanisms involved in coronal heating.
Observations have suggested that it may be uniform \citep[as concluded for
example by][]{pri98}, occur predominantly at the loop top
\citep[e.g.][]{reale02, martens02, schmelz06} or perhaps at the footpoints
\citep[e.g.][]{ant99, asc01, pat04, gudiksen05a}.  However, if the heating
is due to nanoflares, simulations by \citet{pat05} show that present day
observations (including the above papers) using spectroscopic lines emitted
around $1\unit{MK}$ or below can give little insight on the localization of
heating along the loop. This is because the diagnostics are measured during
the cooling of the loop, long after the actual heating has occurred, and so
depend only weakly on the location of the heating.  The original spatial
distribution of the heating has been smoothed out by efficient thermal
conduction along the magnetic field.

Furthermore, the multi-thermal nature of observed loops \citep{schmelz06}
indicates that they are composed of multiple unresolved sub-loops (strands),
although this is controversial.  In this scenario, each strand is heated
intermittently and reaches a high temperatures (perhaps in excess of
$10^7\unit{K}$ in an active region) before being cooled first by conduction
and then by radiation \citep{carg94}.  As heat transport is inefficient
across magnetic field lines (\ie across strands), the whole process in each
strand can usually be considered to be independent from the other strands.

This paper sets out to determine the spatial location of heating in
magnetically closed structures. We simulate a thin coronal loop (\ie a
strand of an observable coronal loop) using a model of reduced-MHD (RMHD)
turbulence and Alfvén waves propagation. Importantly, we take into account
for the first time the effect of a density stratification based on a one
dimensional hydrodynamic loop model. This represents an important first step
in coupling energy transport along a magnetic field with MHD processes. The
role of a non-uniform magnetic field, as well as the effect of the
temperature-dependence of the diffusivity coefficients are discussed.

Our paper is organized as follows: in Sec.~\ref{sec:mod} we describe the
models we use, in Sec.~\ref{sec:simul} we present the numerical simulations
and their results in different cases, and in Sec.~\ref{sec:disc} we discuss
the results further and we give some conclusions.

\section{Description of the model}
\label{sec:mod}

The main model used in this paper is the \shellatm\ model introduced in
\citet{buchlin07pre}, which models MHD turbulence in a coronal loop,
determining the velocity and magnetic field amplitudes as a function of
time, position along the loop and wavenumber $\kperp$ in the transverse
direction.  The shell model uses a mass density profile produced by an
equilibrium run of the \hydrad\ model which calculates the temperature and
density along the magnetic field \citep{bradshaw03a}.  Both models are now
briefly described.

\subsection{The \shellatm\ model}
\label{sec:shellatm}

The \shellatm\ code\footnote{This code is publicly available from\\
  \url{http://www.arcetri.astro.it/~eric/shell-atm/codedoc/}}
\citep{buchlin07pre} is a model of MHD turbulence in a flux tube permeated
by a strong magnetic mean field $\Bo$.  It is based upon reduced-MHD
\citep[RMHD:][]{str76}; in this approximation of incompressible,
low-$\beta$, single fluid MHD, the magnetic field is decomposed into $\vec B
= \Bo + \vec B_\perp$, where $\Bo$ defines the $z$-direction (axial, or
parallel direction) and $B_\perp$ is a small perpendicular fluctuation
($B_\perp \ll B_\parallel$); the velocity field is also a small
perpendicular fluctuation: $\vec u = \vec u_\perp$ with $u_\perp \ll
B_\parallel / \sqrt{\mu_0 \rho}$, where $\rho$ is the mass density.

A further simplification is made, which allows us to perform simulations
that are much longer and with much higher Reynolds numbers than direct
numerical simulations of the RMHD equations \citep[e.g.][]{dmi03b}; this is
needed due to the very wide range of scales involved in coronal turbulence
and because we need to perform long runs in order to get average profiles of
the heating.  The simplification consists of modelling the non-linear
dynamics of RMHD in planes perpendicular to $\Bo$ using a shell-model
\citep{giu98}. Fourier space (which corresponds to a perpendicular section
of the RMHD model at position $z$ and of width $\ell(z)$) is divided into
exponentially-spaced concentric shells $k_n(z) = (2\pi/\ell(z)) 2^n$, each
containing a scalar value for the velocity and magnetic field fluctuations
perpendicular to $B_\parallel$ defined as $u_n(z)$ and $b_n(z)$, where
$b(z)$ has been normalized by $\sqrt{\mu_0 \rho(z)}$. It is assumed that
non-linear interactions are only allowed between triads (triplets) of
neighboring shells.

All of the shell-models distributed along $\Bo$ are coupled through the
Alfvén waves as described by the RMHD equations.  The wave propagation takes
into account the effects of a possible non-uniformity of the profiles of the
mass density $\rho(z)$ and of the axial magnetic field\footnote{Please note
  that giving a profile of $B_\parallel(z)$ is equivalent to giving a
  profile of the flux tube width $\ell(z)$, due to the conservation of the
  magnetic flux in the flux tube.} $B_\parallel(z)$ or of the Alfvén speed
$b_\parallel(z) = B_\parallel(z) / \sqrt{\mu_0 \rho(z)}$, according to
\citet{velli93}.  The model equations are:
\begin{equation}
  \label{eq:dtz}
  (\partial_t \pm b_\parallel \partial_z) Z_n^\pm 
  \pm \frac14 Z_n^\pm \partial_z(\ln\rho)
  \pm \frac12 Z_n^\mp \partial_z b_\parallel
  = -k_n^2 (\nu^+ Z_n^\pm + \nu^- Z_n^\mp) + i k_n
  T_n^{\pm *}
\end{equation}
where $Z_n^\pm(z) = u_n(z) \pm b_n(z)$ are Elsässer variables, $\nu^\pm =
\nu \pm \eta$, $\nu$ is the kinematic viscosity and $\eta$ is the magnetic
diffusivity, and the non-linear terms $T_n^\pm$ are given by Eq.~(4) of
\citet{giu98}
(with $\lambda=2$, $\alpha=2$, $\delta = 5/4$ and $\delta_m = -1/3$).

\begin{figure}[tp]
  \centering
  \includegraphics[width=\linewidth]{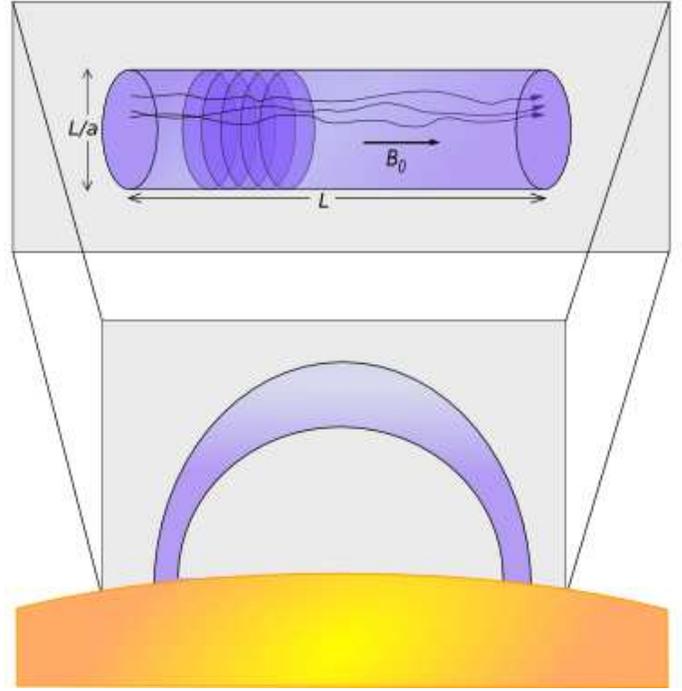}
  \caption{Layout of the \shellatm\ model in the case of a coronal loop:
    shell-models are piled up along $\Bo$ (top) and represent a loop, whose
    footpoints are anchored in the photosphere (bottom).  Reproduced
      from \citet{buchlin07pre}.}
  \label{fig:loop}
\end{figure}

We apply this model to a coronal loop.  As shown in Fig.~\ref{fig:loop}, the
perpendicular planes (each containing a shell-model) represent
cross-sections of the loop; the two end planes represent the photospheric
footpoints of the loop; the large-scale velocity field $u_f$ at the
footpoints is chosen to mimic the photospheric motions:
\begin{equation}
  \label{eq:shellforc}
  u_{z, n}(t) = u_{f,n} \left(
    \Exp{2 i \pi A_{z, n}} \sin^2 (\pi t / t^*) +
    \Exp{2 i \pi B_{z, n}} \sin^2 (\pi t / t^* + \pi / 2)
  \right)
\end{equation}
where $z$ is $0$ or $L$ (both footpoints), $n$ corresponds the the scales
$2\pi/k_n$ of the supergranulation, $A$ and $B$ are random numbers chosen as
detailed in \citet{buchlin07pre}.  This forcing is what injects the energy
into the model, and its amplitude is of the order of $2\unit{km/s}$.

The model gives the energy per unit volume as $E_V=\frac12 \rho \sum_n
(|u_n|^2 + |b_n|^2)$ and the heating (power of energy dissipation) per unit
volume is calculated by computing the time derivative of $E_V$.  This gives
$H_V = \rho \sum_n k_n^2 (\nu |u_n|^2 + \eta |b_n|^2)$ (where all quantities
may depend on $z$): the heating per unit length is $H_L = H_V \pi \ell^2 /
4$ and the heating per unit mass is $H_M = H_V / \rho$ (again, all
quantities may depend on $z$).

\subsection{The \hydrad\ model}
\label{sec:hydrad}

The \hydrad\ code \citep{bradshaw03a} is a hydrodynamic model of a coronal
loop with a self-consistent treatment of radiation (including
non-equilibrium ion populations).  In this paper it is simply used in an
equilibrium run, so as to get a hydrostatic profile
$\rho_{\textsc{HydRad}}(z)$ of the mass density along the loop.  We used an
isothermal hydrostatic solution so that the density along the loop is
stratified due to gravity alone.  Since there is no thermal flux (isothermal
temperature profile) the background heating and radiation are in balance,
and are thus identically zero.  Therefore, we make no assumptions about the
nature of the background heating; once coupled to the \shellatm\ model, all
of the energy released into the loop will be from the shell model only.  The
loop length used in this run was $10\unit{Mm}$.

\subsection{A simple phenomenology for the energy and heating profiles}
\label{sec:phen}

In this section we discuss what profiles of the fields, energy and
dissipation power along the loop would be expected from a simple
phenomenology.

A mass density profile $\rho(z)$, a magnetic field profile $B_\parallel(z)$
and a profile $\nu(z)$ of the dissipation coefficient (kinetic or magnetic)
are given (in model units).  The Elsässer fields $Z^\pm$ are assumed to be
both of the same order of magnitude $Z$ (which is reasonable in a loop).
The effects of energy dissipation and of non-linear interactions on the
magnitudes of the fields are assumed to be negligible.

The conservation of energy fluxes $\phi_E^\pm = \pm b_\parallel E_V^\pm
\pi\ell^2 / 4$ together with the conservation of the magnetic flux $\phi_B =
B_\parallel \pi \ell^2 / 4$ gives then a profile $E_V \propto \rho^{1/2}$ of
the energy per unit volume.  The dissipation power per unit volume is $H_V =
\rho \nu \kperp^2 Z^2 \propto \nu \kperp^2 \rho^{1/2}$ for a fixed $\kperp$.
However, one must bear in mind that when a wave travels along the loop (with
non-linear interactions neglected), the associated $\kperp$ is affected by
the variation of the width of the loop (in the model, the wave stays in the
same shell $n$ of the model): $\kperp = k_n(z) = k_0(z) \lambda^n$, with
$k_0 \propto B_\parallel^{1/2}$ because of the conservation of magnetic
flux.  For this reason, for a linear wave in the shell $n$, $H_V^{(n)}
\propto \nu B_\parallel \rho^{1/2}$.

\section{Numerical simulations}

\label{sec:simul}

For all simulations, we choose model units so that the length of the loop is
$L = 1$, and so as to have the below-mentioned values for the magnetic field
$B_{\parallel}^f$ and for the mass density $\rho_f$ at the footpoints.
Conversion factors to physical units for a sample loop are given in
Table~\ref{tab:conv}, but to use the same simulations for other loops with
different physical parameters it is sufficient to take different values for
the basic conversion factors $M$, $L$ and $T$ for mass, distance and time
respectively.  We stress that these conversion factors do not necessarily
represent typical values of the actual physical quantities in the loop (the
model variables are not necessarily of the order of unity in model units;
this will especially be true for the diffusion coefficients).

If $\rho_{\textsc{HydRad}}(z)$ is the density profile from the equilibrium
run of the \hydrad\ code (see Sec.~\ref{sec:hydrad}), we define $R(z) =
\rho_{\textsc{HydRad}}(z) / \rho_{\textsc{HydRad}}(0)$, with $0$ being the
position of one footpoint.  $R$ varies between $R_t = 1/30$ at the loop top
and $R_f = 1$ at the footpoints. The profiles of mass density and Alfvén
speed are then chosen as a function of $R$ as specified in
Table~\ref{tab:par} for each of the runs.  The physical conditions are
chosen to be the same at the footpoints in each run ($\rho_f = 1 / R_t$ and
$B_{\parallel}^f = 1$), so that the boundary conditions of the model remain
the same in the different runs.

A typical value for the diffusivity coefficients $\nu$ and $\eta$ is
$5\,10^{-10}$ in model units (see Table~\ref{tab:par} for the exact values
used in each run), \ie of the order of $5\,10^3\unit{m^2/s}$; typical
Reynolds and magnetic Reynolds numbers are $10^6$.  This is lower than in
the real corona (diffusivity coefficients are higher than in the real
corona) but it represents a huge improvement over direct numerical
simulations of MHD.

\begin{table}[tp]
  \centering
  \caption{Conversion factors between model units and S.I.\ physical units:
    expression in the general case and value for a typical modelled
    loop.}
  \label{tab:conv}
  \begin{tabular}{lcc}
    \hline \hline
             & \multicolumn{2}{c}{Conversion factor to S.I.\ physical units} \\
    Quantity & Expression & Typical value \\ \hline
    Mass & $M$ & $10^9\unit{kg}$ \\ 
    Length, distance & $L$ & $10^7\unit{m}$ \\
    Time & $T$ & $10\unit{s}$ \\\hline
    Velocity, Alfvén speed & $L T^{-1}$ & $10^6\unit{m\,s^{-1}}$ \\
    Magnetic field & $\sqrt{\mu_0} M^{1/2} L^{-1/2} T^{-1}$ &
    $1.1\,10^{-3}\unit{T}$\\
    Mass density & $M L^{-3}$ & $10^{-12}\unit{kg\,m^{-3}}$ \\
    Diffusivity & $L^2 T^{-1}$ & $10^{13}\unit{m^2\,s^{-1}}$ \\
    Energy per unit volume & $M L^{-1} T^{-2}$ & $1\unit{J\,
      m^{-3}}$ \\
    Power per unit volume & $M L^{-1} T^{-3}$ &
    $10^{-1}\unit{W\,m^{-3}}$\\ \hline
  \end{tabular}
\end{table}

We use the model with $1000$ planes piled up along $\Bo$, and each of these
has $18$ shells ($n = 0, \ldots 17$).  Each of the simulations is run during
$500$ units of time (after an initial phase of energy growth), which
corresponds to the order of 100 large-eddy turn-over times.  Profiles of the
energy and of the dissipation per unit volume are averaged over the duration
of the runs.

\newcommand*\runu{(u)}
\newcommand*\runa{(a)}
\newcommand*\runb{(b)}
\newcommand*\runc{(c)}
\newcommand*\runA{(A)}
\newcommand*\runB{(B)}
\newcommand*\runC{(C)}
\newcommand*\runp{(p)}
\newcommand*\runq{(q)}
\newcommand*\runr{(r)}
\newcommand*\runs{(s)}
\newcommand*\runt{(t)}

\begin{table}[tp]
  \centering
  \caption{Summary of the parameters of the modelled loop for different
    runs, in model units.}
  \label{tab:par}
  \begin{tabular}{llllll}
    \hline \hline
    Run & $\rho R_t$ & $b_\parallel / R_t^{1/2}$ & $B_\parallel$ & $\nu / 10^{-10}$ & $\eta / 10^{-10}$ \\
    \hline 
    \runu & $1$       & $1$        & $1$         & $5$                   & $5$                   \\[.8ex]
    \runa & $R$       & $R^{-1/2}$  & $1$        & $5$                   & $5$                   \\
    \runb & $R^{1/2}$ & $R^{-1/4}$  & $1$        & $5$                   & $5$                   \\
    \runc & $R^{1/4}$ & $R^{-1/8}$  & $1$        & $5$                   & $5$                   \\[.8ex]
    \runA & $R$       & $R^{-1/4}$  & $R^{1/4}$  & $5$                   & $5$                   \\
    \runB & $R^{1/2}$ & $R^{-1/8}$  & $R^{1/8}$  & $5$                   & $5$                   \\
    \runC & $R^{1/4}$ & $R^{-1/16}$ & $R^{1/16}$ & $5$                   & $5$                   \\[.8ex]
    \runp & $R$      & $R^{-1/2}$   & $1$        & $0.5 (R/R_t)^{-7/2}$  & $0.5 (R/R_t)^{3/2}$    \\
    \runq & $R$       & $R^{-1/2}$  & $1$        & $50 (R/R_t)^{-7/2}$   & $0.5 (R/R_t)^{3/2}$    \\
    \runr & $R$       & $R^{-1/2}$  & $1$        & $10^2 (R/R_t)^{-7/2}$ & $10^{-2}(R/R_t)^{3/2}$ \\
    \runs & $R$       & $R^{-1/2}$  & $1$        & $10^2 (R/R_t)^{-7/2}$ & $10^{-4}(R/R_t)^{3/2}$ \\
    \runt & $R$       & $R^{-1/2}$  & $1$        & $10^2 (R/R_t)^{-7/2}$ & $10^{-6}(R/R_t)^{3/2}$ \\
    \hline
  \end{tabular}
\end{table}

\subsection{Test case: uniform loop}
\label{sec:testcase}

\begin{figure}[tp]
  \centering
  \includegraphics[width=\linewidth]{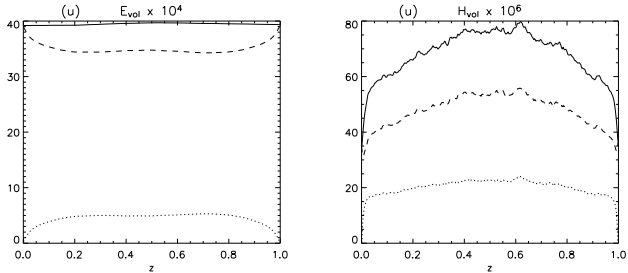}
  \caption{Average profiles of the energy (left) and dissipation power
    (right) per unit volume, as a function of the position $z$ along the
    loop, for run \runu.  The dotted lines are the kinetic energy or
    dissipation power, the dashed line the magnetic energy or dissipation
    power, and the plain line is the total.}
  \label{fig:resu}
\end{figure}

We present run \runu\ as a test case, where the mass density, magnetic
field, Alfvén speed and diffusion coefficients are all uniform (as in
\citealt{buchlin07pre}).  Figure~\ref{fig:resu} shows that the time-average
profile of the total energy per unit volume is uniform along the loop, with
the magnetic energy dominating.  The kinetic energy density is also uniform
except near the footpoints, where it gets lower because of the boundary
conditions ($u_f \ll \Bo$).

The energy dissipation (right panel of Fig.~\ref{fig:resu}) is slightly
non-uniform: although the phenomenology of Sec.~\ref{sec:phen} suggests a
uniform profile, there is a drop near the footpoints and an enhancement near
the loop top; the former can be attributed mainly to the contribution of the
profile of kinetic energy dissipation, and the latter to the profile of the
magnetic energy dissipation.  The ratio of total dissipation power in the
central $10\%$ of the loop length to the total dissipation power in both
$5\%$ of the loop length at the footpoints is $1.5$.

\begin{figure}[tp]
  \centering
  \includegraphics[width=\linewidth]{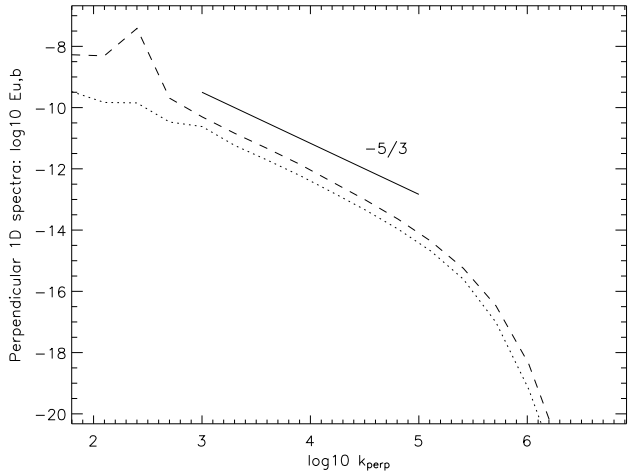}
  \caption{Time-averaged kinetic (dots) and magnetic (dashes)
    perpendicular 1D energy spectra in the loop for run \runu.  A line of
    slope $-5/3$ is shown for reference.  These spectra are representative
    of the spectra obtained for the other runs, although there can be some
    variations between runs.}
   \label{fig:uspec}
\end{figure}

A second observation is that the ratio of magnetic to kinetic energy
dissipation is lower than the ratio of the magnetic to kinetic energy.  This
is because most magnetic energy is concentrated at large
scales\footnote{This is a well-known effect coming from the resonances in
  the solution of the linearized RMHD equations \citep{mil97}.} as can be
seen on the spectra of Fig.~\ref{fig:uspec}, which do not dissipate as much
as the smaller scales.  Then the relative weight of the kinetic dissipation
in the total dissipation is increased compared to the weight of the kinetic
energy in the total energy; as the kinetic dissipation profile is lower at
the footpoints (because of the boundary condition with $u_f \ll \Bo$), this
explains the non-uniform shape of the total dissipation profile.  This kind
of effect needs to be kept in mind when analyzing the other runs.

\subsection{Loop with a non-uniform mass density}
\label{sec:loopa}

\begin{figure}[tp]
  \centering
  \includegraphics[width=\linewidth]{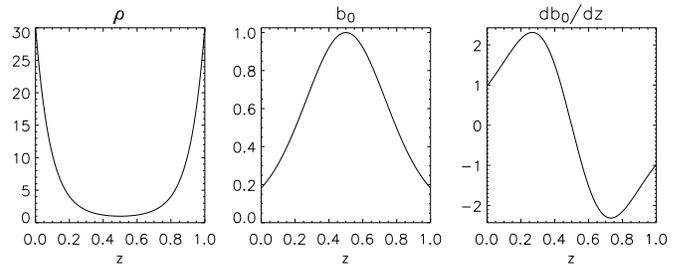}
  \caption{Profiles of mass density $\rho$, Alfvén speed $b_\parallel$ and
    gradient of Alfvén speed $\ud b_\parallel / \ud z$ for run \runa, as a
    function of the position $z$ along the loop.}
  \label{fig:arhova}
\end{figure}

We now consider a more realistic loop, with a non-uniform mass density
determined by the \hydrad\ model.  The ability to take into account a
non-uniform mass density (as well as a variable axial magnetic field: see
later) is an important advantage of the \shellatm\ model when compared to
the model of \citet{nig04} as used by \citet{reale05}.  Run \runa\ is
performed using a mass density profile $\rho(z)=R(z) / R_t$
(Fig.~\ref{fig:arhova}). In runs \runb\ and \runc\ we use $R^{1/2} / R_t$
and $R^{1/4} / R_t$ for the mass density profiles, so as to assess the
effect of smoother profiles.

Loops seen with Yohkoh and TRACE have only a slight width enhancement at the
top relative to the footpoints \citep{klimchuk92, klimchuk00, watko00}.
It is thus reasonable to assume a uniform axial
magnetic field along such loop. In the model we set $B_{\parallel} =
{b_{\parallel}}^t \sqrt{\rho_t}$ with ${b_{\parallel}}^t=1$ ($\mu_0=1$ in model
units). The Alfvén speed is then $b_\parallel(z) = {b_{\parallel}}^t (\rho(z)
/ \rho_t)^{1/2}$, a function of $z$, as shown in Fig.~\ref{fig:arhova}.
Consequently the waves are partially reflected due to the Alfvén speed
gradient .

\begin{figure}[tp]
  \centering
  \includegraphics[width=\linewidth]{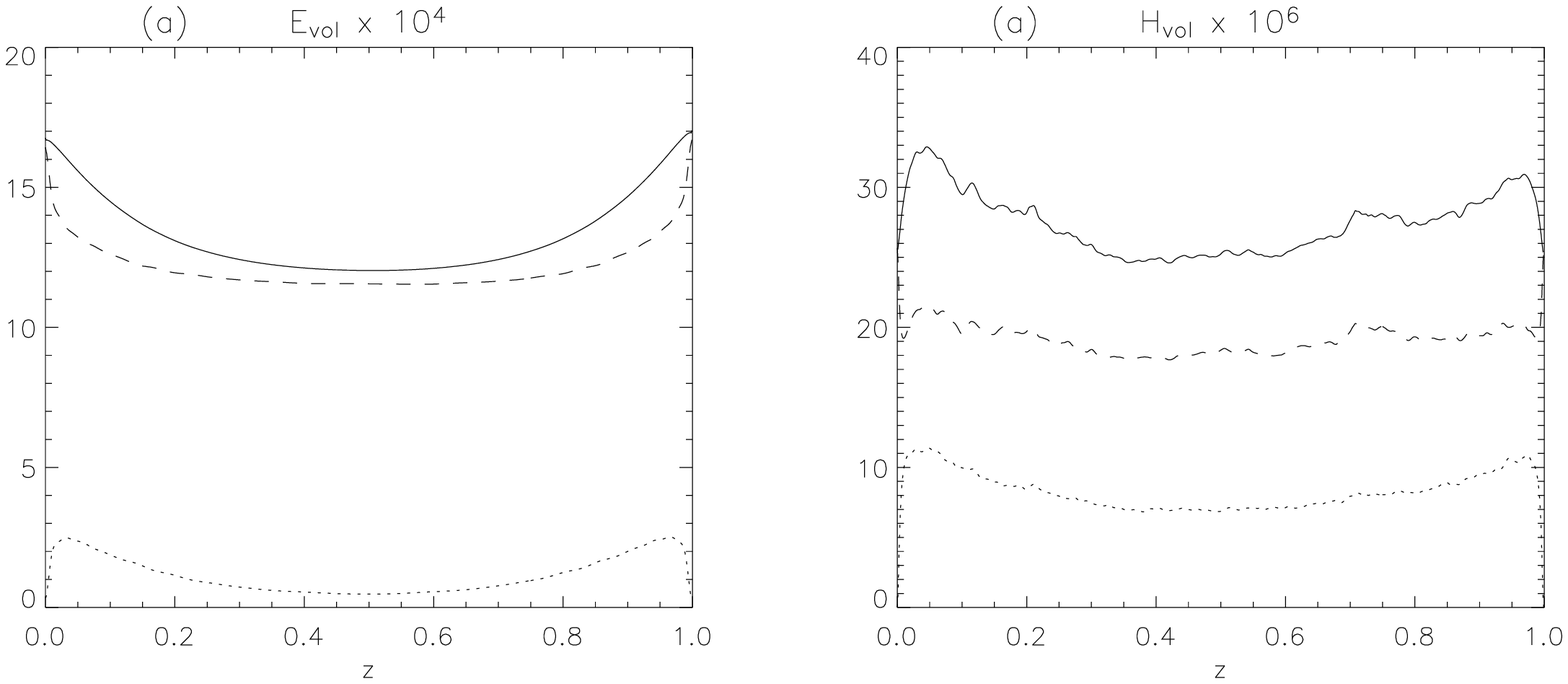}
  \includegraphics[width=\linewidth]{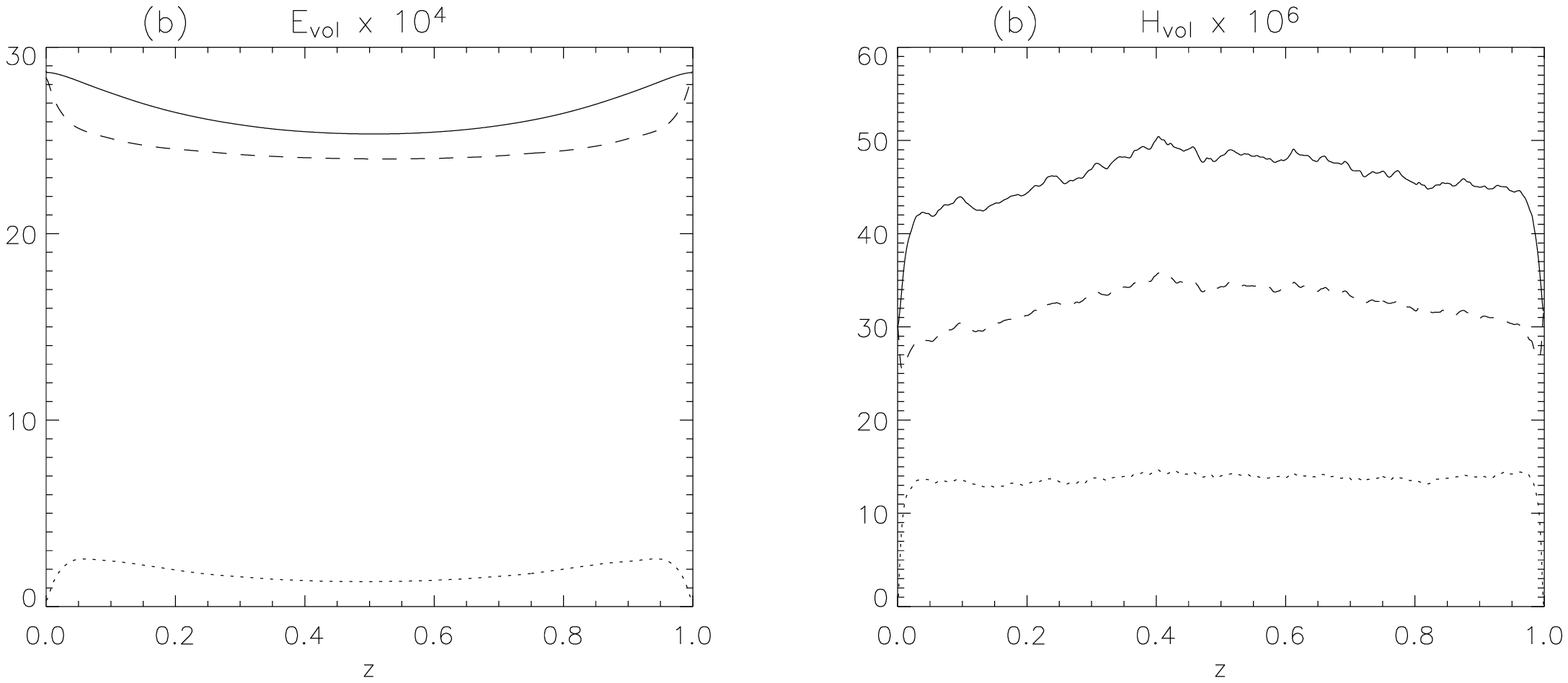}
  \includegraphics[width=\linewidth]{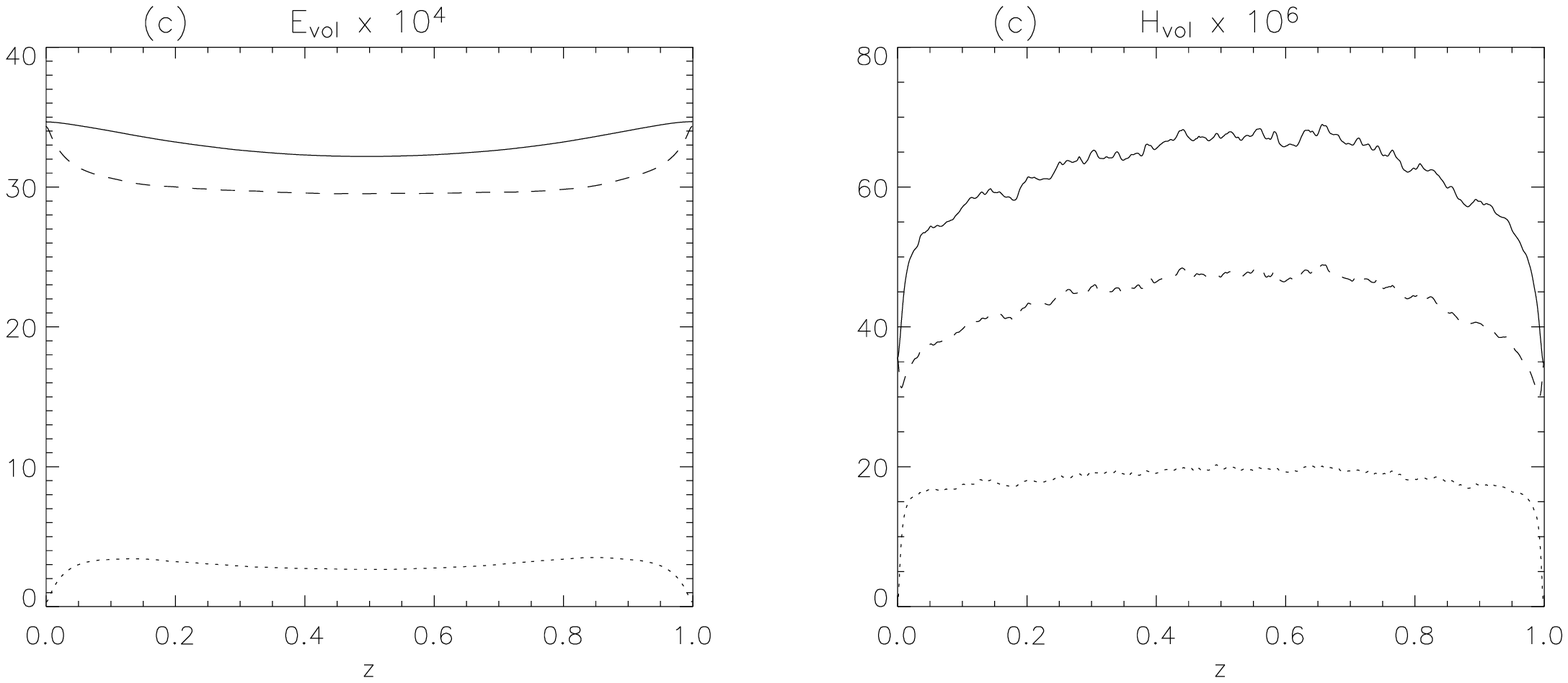}
  \caption{Average profiles of the energy (left) and dissipation power
    (right) per unit volume, as a function of the position $z$ along the
    loop, for runs \runa, \runb\ and \runc\ (from top to bottom).  The
    dotted lines are the kinetic energy or dissipation power, the dashed
    line the magnetic energy or dissipation power, and the plain line is the
    total (same as in Fig.~\ref{fig:resu}).}
  \label{fig:resa}
\end{figure}

The resulting time-averaged profiles of the energy per unit volume $E_V(z)$
and of the heating per unit volume $H_V(z)$ are shown in
Fig.~\ref{fig:resa}.  The striking result is that in all these cases,
including run \runa\ which has steep gradients of density and Alfvén speed,
the effect of the non-uniform density on the heating profiles is very
limited.  For run \runa\ for example, while the phenomenology of
Sec.~\ref{sec:phen} predicts a dependence in $\rho^{1/2}$ for both energy
density and dissipation, \ie a ratio $\approx 5.5$ between the footpoints
and the loop top, the actual ratio is only about $1.3$ between the highest
values (near the footpoints) and the lowest values, for both energy and and
dissipation.  For runs \runb\ and \runc\ (which have shallower density
gradients), the profiles are even closer to being uniform.

We performed fits of each of these profiles to a power-law $\rho^\alpha$ of
the mass density (Table~\ref{tab:expind}).  The domain for the fit
was restricted to the central $90\%$ of the loop, to avoid effects due to
the boundary conditions.  In cases \runa\ and \runb, the fit of the heating
profile gives an index $\alpha_H$ close to $0$ for the power-law
$\rho^{\alpha_H}$ (taking into account the error bars of the fit; see
Table~\ref{tab:expind} for the detailed results of the fits).  Each of the
individual profiles cannot be considered as uniform (as would be the case if
the fit was perfect and with $\alpha_H=0$), but they are still considerably
more uniform than that expected from the steep gradient of the mass density
$\rho$.

The fit of the energy profiles gives an index $7.1\,10^{-2} < \alpha_E <
9.4\,10^{-2}$ (in all three cases, taking into account the error bars).  The
profile of energy is thus also much more uniform than expected from the
phenomenology ($\alpha = 1/2$).  This is surprising as it means that, when
going from the footpoints to the loop top, the energy decreases slower than
what would be expected from the conservation of the energy flux (under the
assumptions detailed in Sec.~\ref{sec:phen}).  The discrepancies between the
phenomenology and the results are discussed further in Sec.~\ref{sec:disc}.

\begin{table}[tp]
  \centering
  \caption{Summary of the indices $\alpha_E$ and $\alpha_H$ for the
    power-law fits $E_V \propto \rho^{\alpha_E}$ and $H_V \propto
    \rho^{\alpha_H}$, (1) expected from the phenomenology of
    Sec.~\ref{sec:phen} and (2) obtained from the numerical simulations, for
    the runs with uniform dissipation coefficients.  The fits from the
    simulations are performed on time-averaged profiles and on the central
    $90\%$ of the loop.} 
  \label{tab:expind}
  \begin{tabular}{lcccc}
    \hline \hline
    Run & $\alpha_E$  (1) & $\alpha_E$ (2) & $\alpha_H$ (1) & $\alpha_H$ (2)       \\ \hline
    \runa & $1/2$ & $(8.50 \pm 0.05) 10^{-2}$ & $1/2$ & $( 7.39 \pm 0.09) 10^{-2}$ \\
    \runb & "     & $(7.14 \pm 0.02) 10^{-2}$ & "     & $(-8.98 \pm 0.15) 10^{-2}$ \\
    \runc & "     & $(9.42 \pm 0.01) 10^{-2}$ & "     & $(-2.85 \pm 0.03) 10^{-1}$ \\
    \runA & "     & $(3.15 \pm 0.00) 10^{-1}$ & $3/4$ & $( 4.59 \pm 0.01) 10^{-1}$ \\
    \runB & "     & $(3.18 \pm 0.00) 10^{-1}$ & "     & $( 3.81 \pm 0.01) 10^{-1}$ \\
    \runC & "     & $(3.05 \pm 0.00) 10^{-1}$ & "     & $( 1.96 \pm 0.04) 10^{-1}$ \\
    \hline
  \end{tabular}
\end{table}

Although the comparison of the heating in loops of different lengths is out
of the scope of this paper, we have also performed a simulation of a
$40\unit{Mm}$-long loop.  In this case a uniform chromospheric temperature
yields unrealistically low coronal densities, so the loop has been divided
into a $20\,000\unit{K}$ chromosphere and, from the points where the density
has been divided by $100$ compared to the footpoints, a $1\unit{MK}$ corona.
Although the ratio of footpoint to loop top densities is higher than in the
other simulations presented in this paper, the average heating profile
remains quasi-uniform ($\alpha_H=(3.13 \pm 0.66) 10^{-2}$).  Therefore, when
the heating model \shellatm\ is considered alone in a hydrostatic loop (i.e.
neglecting the feedback effects of the cooling processes), longer loops do
not seem to behave differently than shorter loops.

\subsection{Loop with non-uniform mass density and magnetic field}
\label{sec:loopn}

In very large structures like coronal streamers, it is likely that the
magnetic field gets weaker with altitude.  We model this variation through a
dependence on the mass density: the axial magnetic field of the loop is
chosen to be $B_\parallel(z) = b_\parallel \sqrt{\mu_0 \rho(z)} (\rho(z) /
\rho_t)^{-1/4}$ (the exponent $1/4$ has been chosen so as to minimize both
the gradients of magnetic field and of Alfvén speed for a given density
profile).  The Alfvén speed is then $b_\parallel (\rho(z) / \rho_t)^{1/4}$.

Runs \runA, \runB\ and \runC\ are performed with this magnetic field
profile, and the same density profiles as runs \runa, \runb\ and \runc\
respectively (see Table~\ref{tab:par}).  For run \runA\ for example, the
ratio of the magnetic field between the footpoints and the loop top is then
$2.3$.

\begin{figure}[tp]
  \centering
  \includegraphics[width=\linewidth]{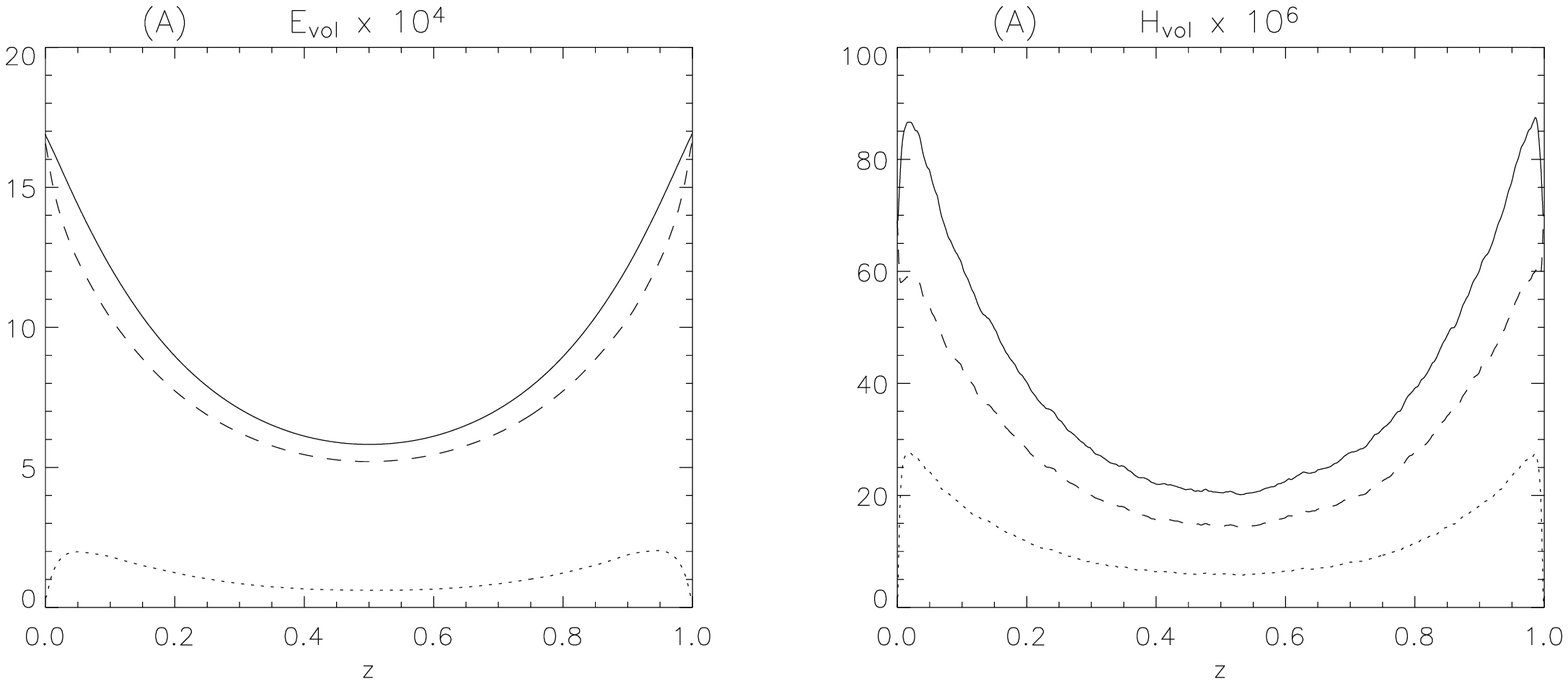}
  \includegraphics[width=\linewidth]{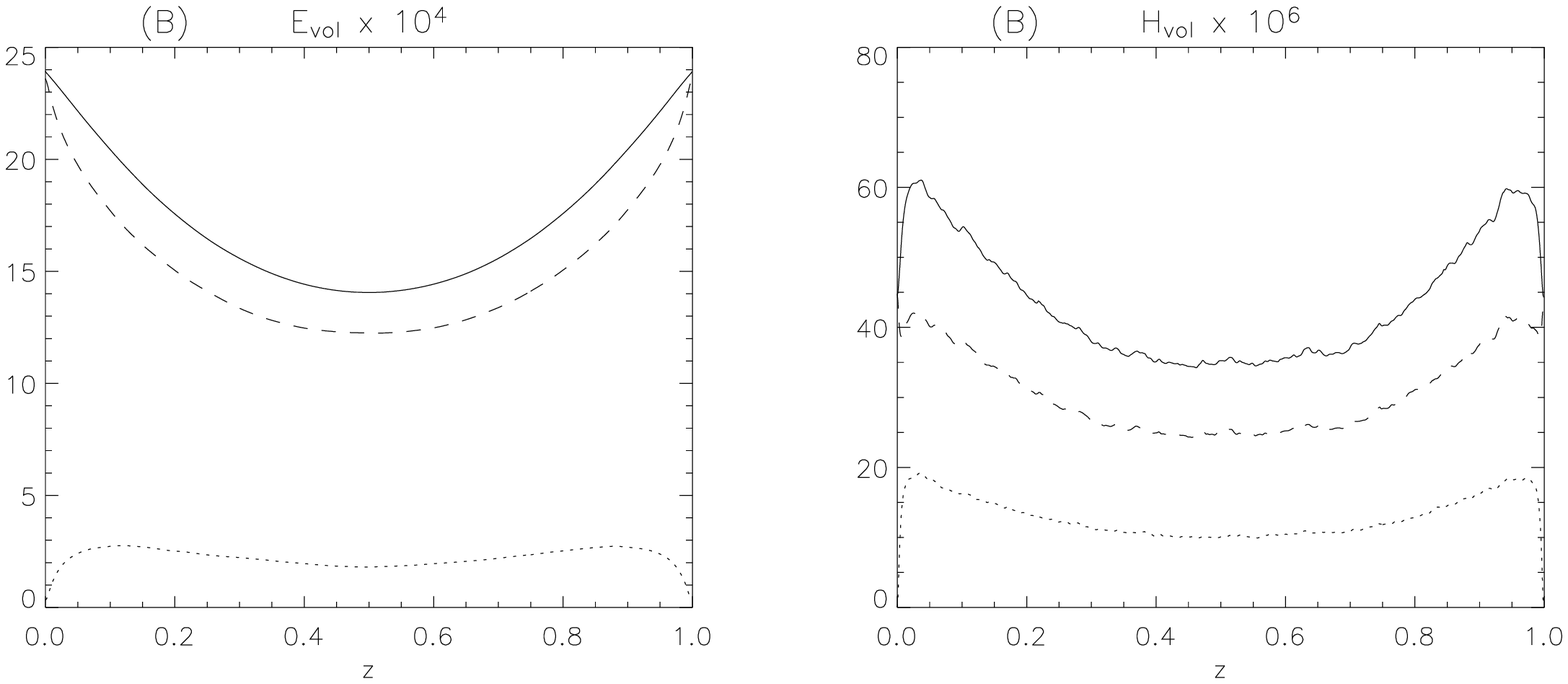}
  \includegraphics[width=\linewidth]{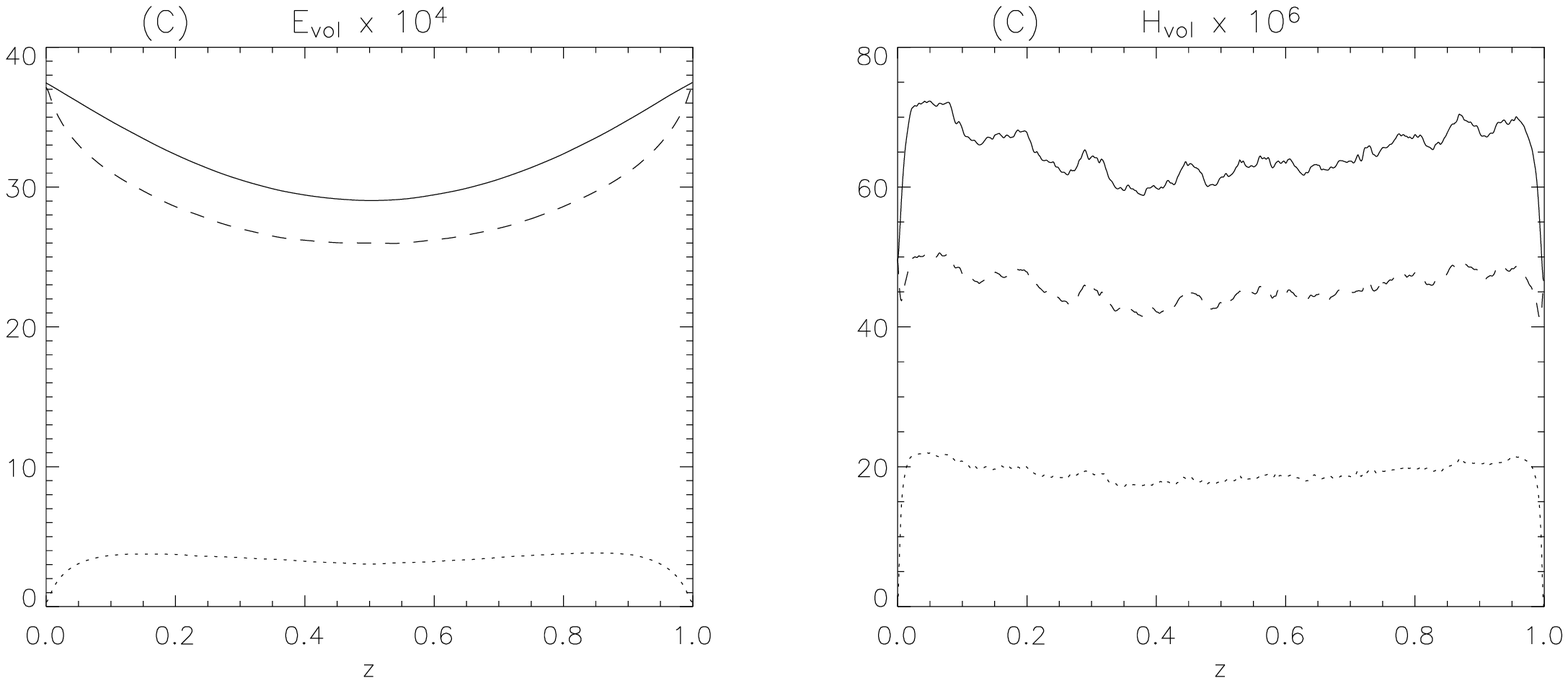}
  \caption{Average profiles of the energy (left) and dissipation power
    (right) per unit volume, as a function of the position $z$ along the
    loop, for runs \runA, \runB\ and \runC\ (from top to bottom).  The
    dotted lines are the kinetic energy or dissipation power, the dashed
    line the magnetic energy or dissipation power, and the plain line is the
    total (same as in Fig.~\ref{fig:resu}).}
  \label{fig:resA}
\end{figure}

The resulting time-averaged profiles of the energy and the heating per unit
volume are shown in Fig.~\ref{fig:resA}.  In all cases (and especially
\runA, which has the steepest density and magnetic field gradients) the
heating is enhanced at the footpoints.  The results of the power-law fits of
the heating profiles as a function of the mass density are also shown in
Table~\ref{tab:expind}; we get $0.19 < \alpha_H < 0.46$ in all three cases
(taking into account the error bars)
whereas the phenomenology of Sec.~\ref{sec:phen} predicts $\alpha_H = 3/4$.

This enhancement of the dissipation power at the footpoints can be seen as a
consequence of the higher energy per unit volume there.  Indeed, the
power-law fits of the energy profiles give $0.30 < \alpha_E < 0.32$ in all
three cases, as seen in Table~\ref{tab:expind} (taking into account the
error bars), whereas the phenomenology predicts $\alpha_E = 1/2$.  Again,
there is more energy at the loop top than what would be expected from the
conservation of the energy flux.

\subsection{Loop with non-uniform mass density and diffusivity coefficients}
\label{sec:vardiff}

Up to now the diffusion coefficients have been assumed to be uniform, but
this is not the case if the temperature profile is not uniform.  So as to
take this effect into account, we consider a different situation than in the
previous sections, with the same density profile $\rho(z) = R(z) / R_t$ as
runs \runa\ and \runA, but assuming now that the loop has a uniform pressure
and that the law of perfect gases holds, hence a variation of a factor $R_f
/ R_t = 30$ of the temperature between the footpoints and the loop top.
With these assumptions, the variations of the diffusivity coefficients in
the magnetized plasma as obtained from \citet{braginskii65} are in $T^{7/2}$
for the kinematic viscosity and in $T^{-3/2}$ for the magnetic diffusivity,
where $T$ is the plasma temperature (which is only known in our model from
the density $\rho$ and the assumptions we make in this section).  We thus
choose profiles $\nu = \nu_t R^{-7/2}$ and $\eta = \eta_t R^{3/2}$ for the
kinematic viscosity and for the magnetic diffusivity respectively, where
$\nu_t$ and $\eta_t$ are the values of $\nu$ and $\eta$ that we have chosen
at the loop top.

\begin{figure}[tp]
  \centering
  \includegraphics[width=\linewidth]{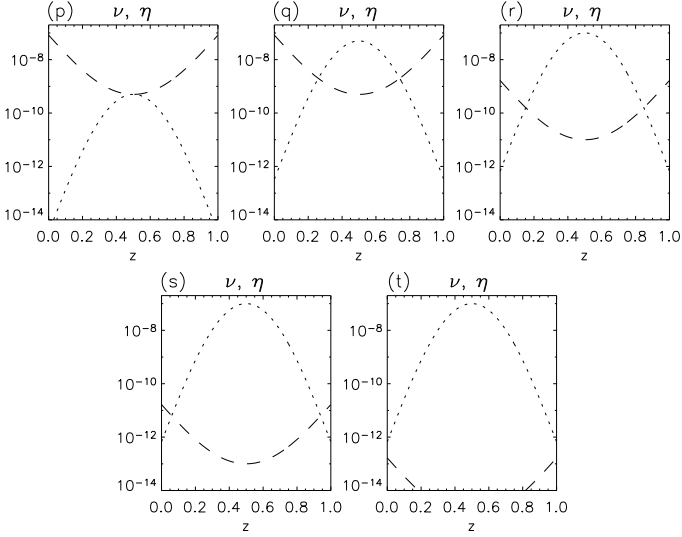}
  \caption{Profiles of the dissipation coefficients $\nu$ (kinematic
    viscosity; dots) and $\eta$ (magnetic diffusivity; dashs) used in runs
    \runp, \runq, \runr, \runs\ and \runt\ (from left to right and top to
    bottom).}
  \label{fig:nup}
\end{figure}

We perform runs \runp, \runq, \runr, \runs\ and \runt\ at different
magnetic Prandtl numbers at loop top: $Pr_M^t = \nu_t / \eta_t = 1$,
$10^2$, $10^4$, $10^6$ and $10^8$ respectively (see Table~\ref{tab:par}).
With the profiles we use, the magnetic Prandtl number is $(R_f/R_t)^5$ times
lower at the footpoints than at the loop top, \ie $2.4\,10^7$ times lower:
it is $Pr_M^f = 4\,10^{-8}$, $4\,10^{-6}$, $4\,10^{-4}$, $4\,10^{-2}$ and
$4$ for runs \runp, \runq, \runr, \runs\ and \runt\ respectively (see
Fig.~\ref{fig:nup}).  These ranges of magnetic Prandtl numbers and of
diffusivity coefficients, which are impossible to explore with direct
numerical simulations, are easily managed by the \shellatm\ model.

\begin{figure}[tp]
  \centering
  \includegraphics[width=\linewidth]{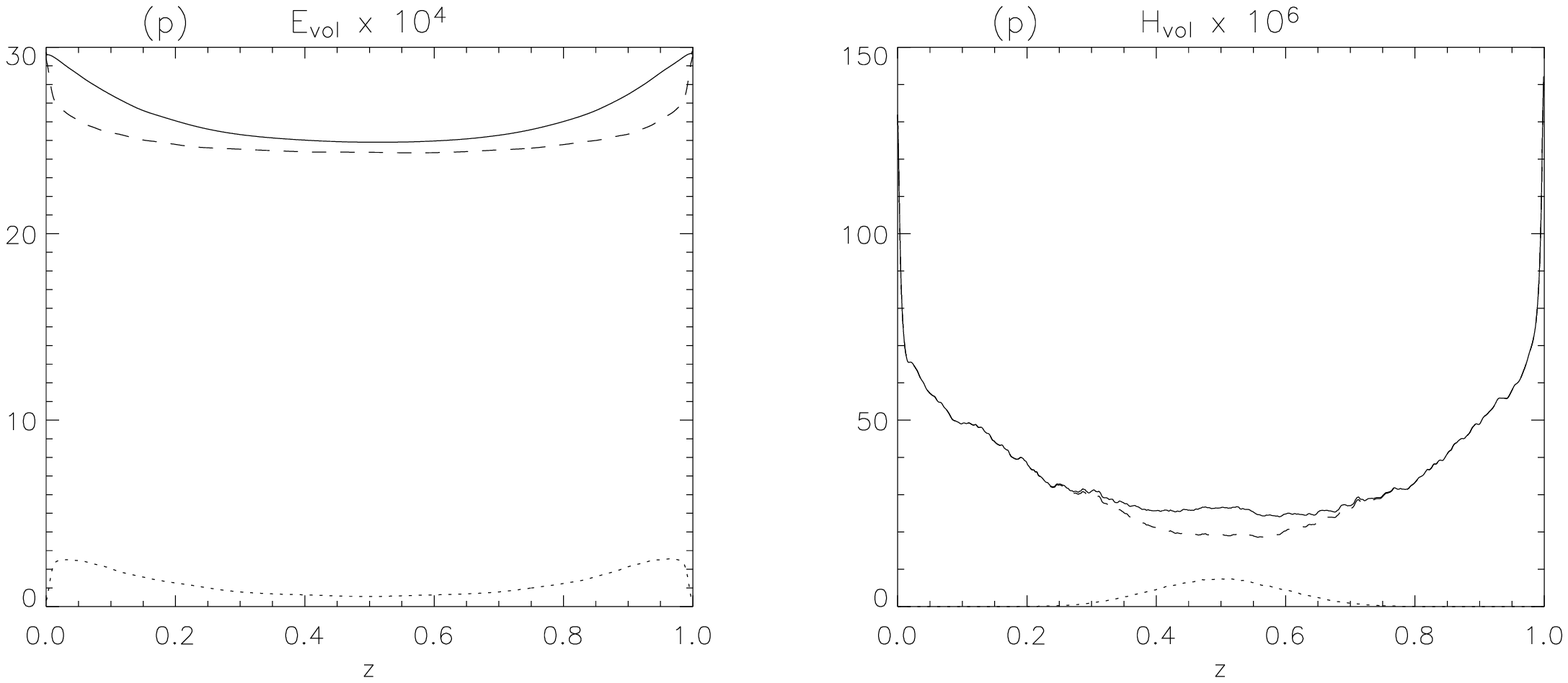}
  \includegraphics[width=\linewidth]{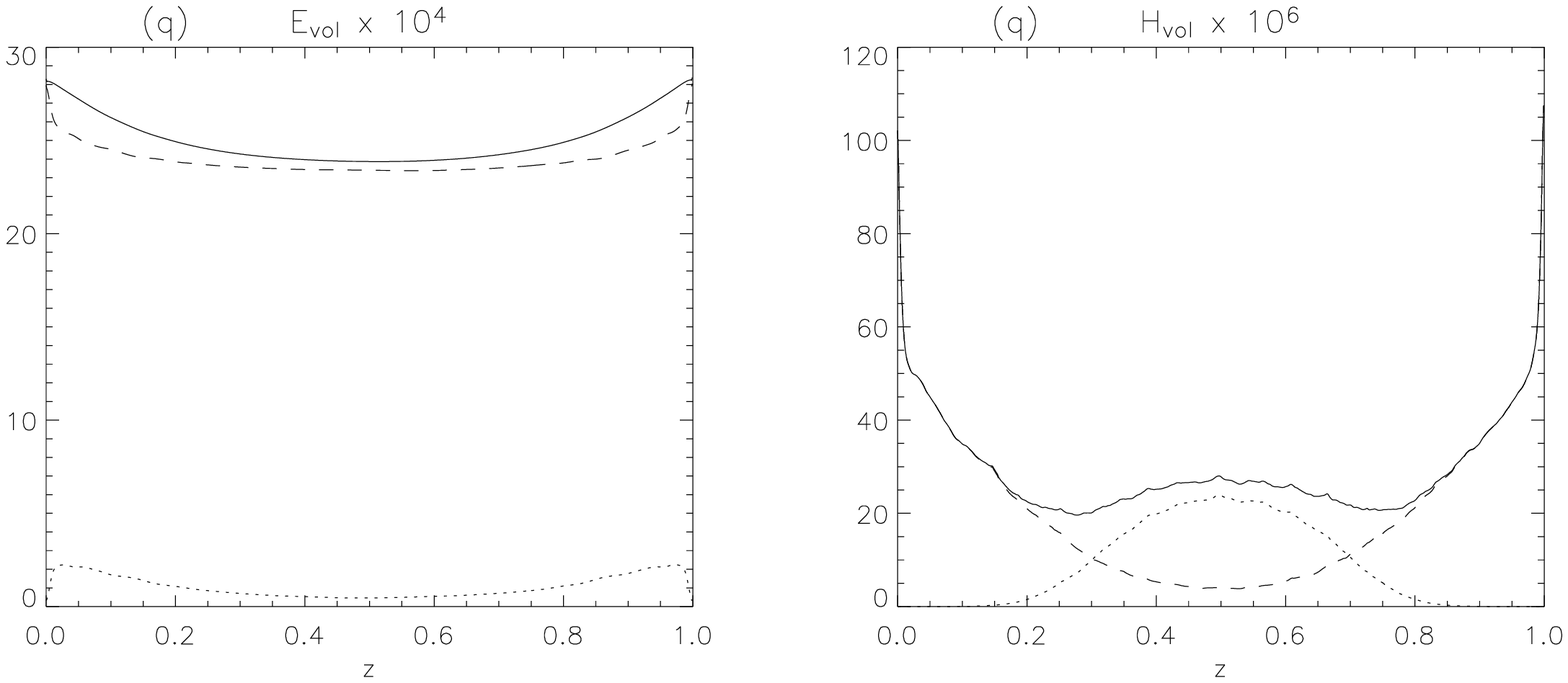}
  \includegraphics[width=\linewidth]{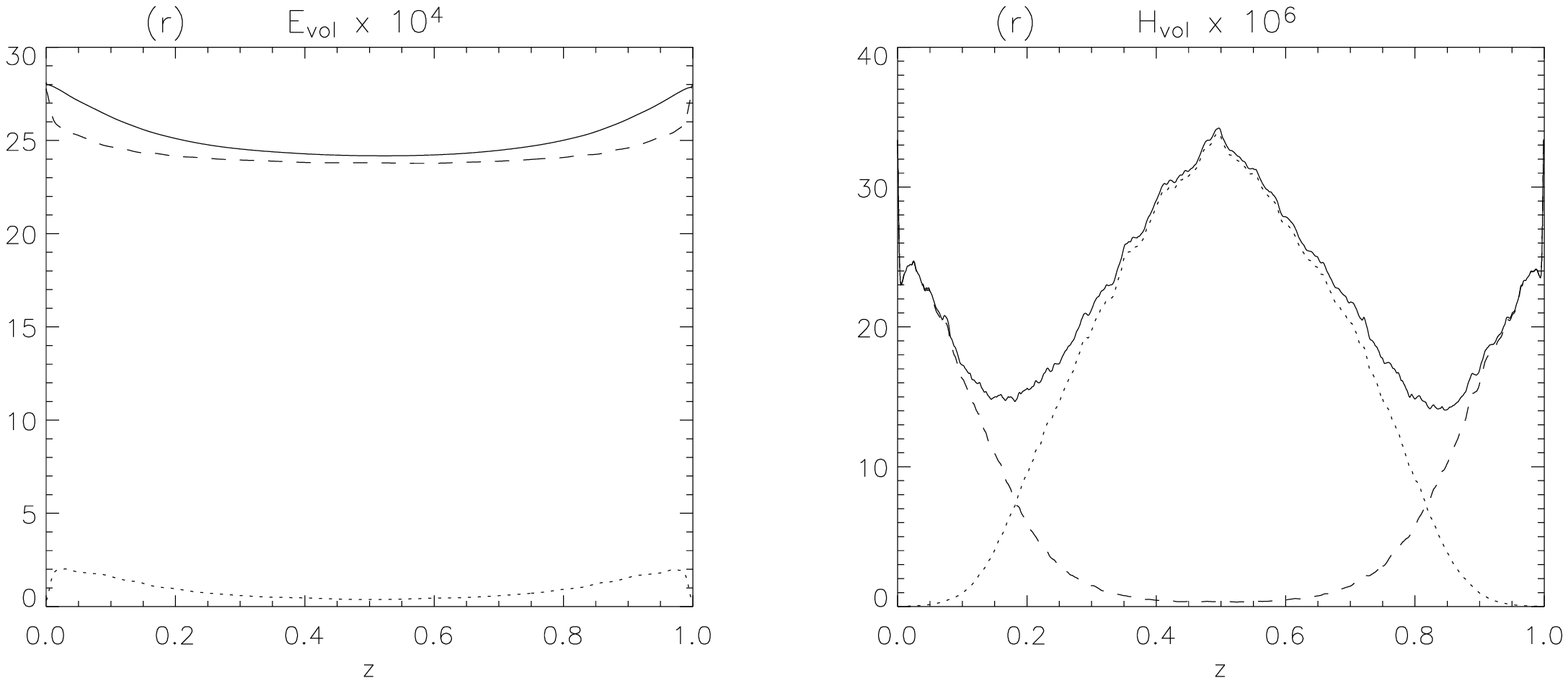}
  \includegraphics[width=\linewidth]{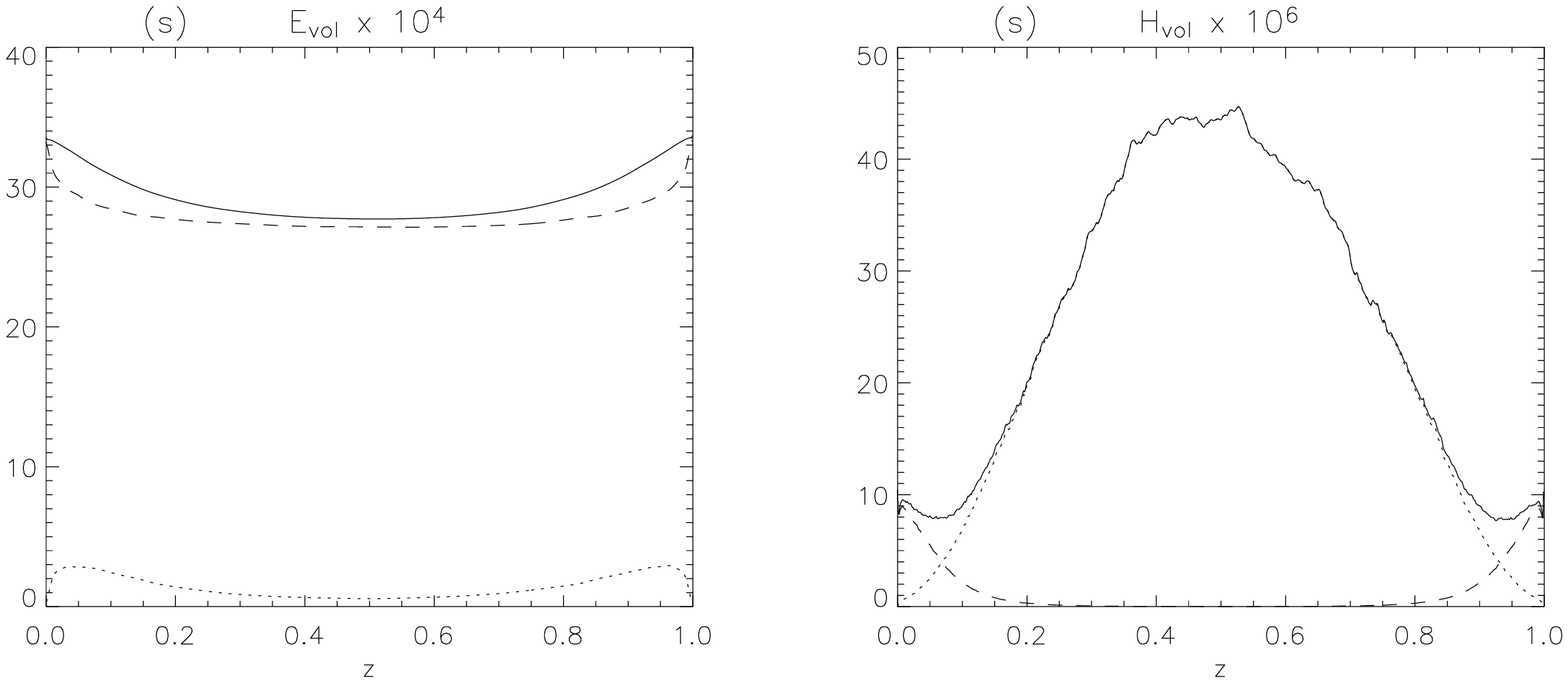}
  \includegraphics[width=\linewidth]{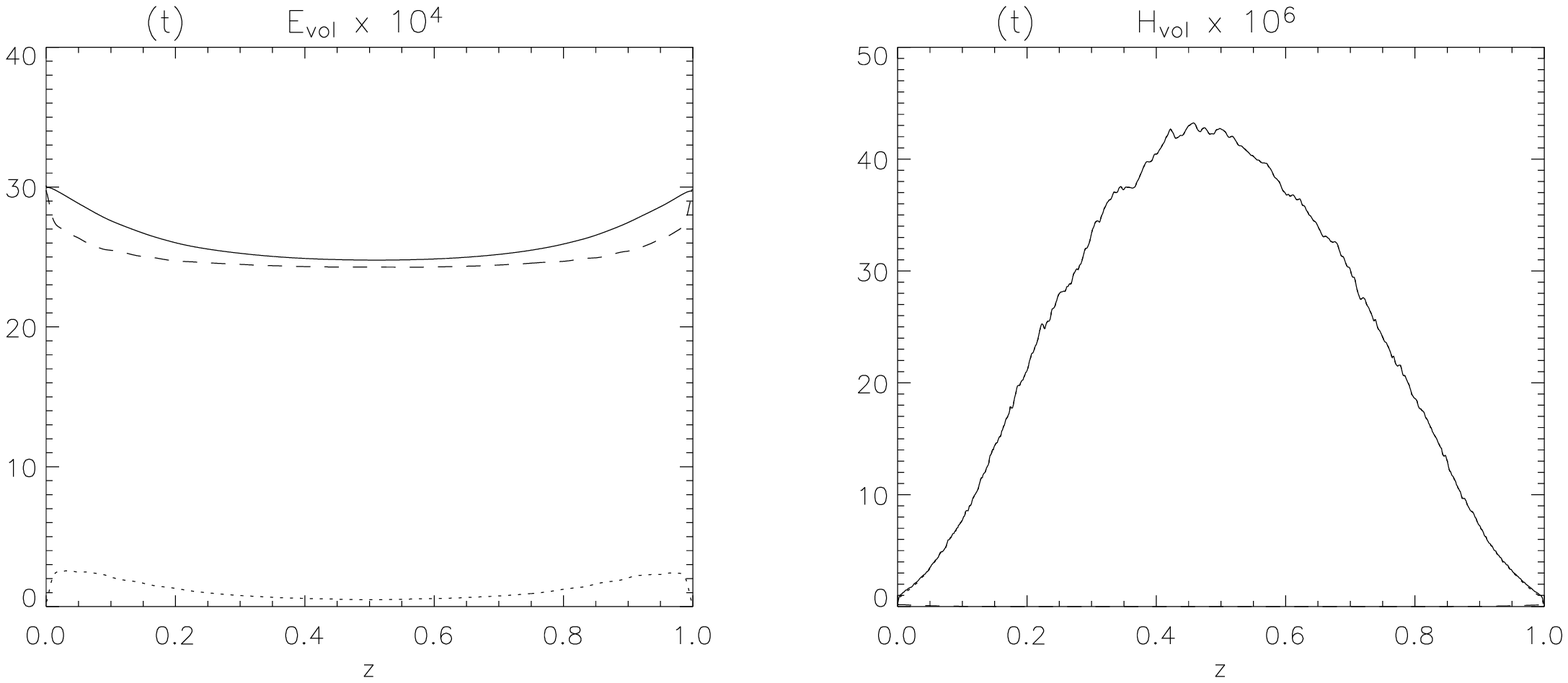}
  \caption{Average profiles of the energy (left) and dissipation power
    (right) per unit volume, as a function of the position $z$ along the
    loop, for runs \runp, \runq, \runr, \runs\ and \runt\ (from top to
    bottom).  The dotted lines are the kinetic energy or dissipation power,
    the dashed line the magnetic energy or dissipation power, and the plain
    line is the total (same as in Fig.~\ref{fig:resu}).}
  \label{fig:resp}
\end{figure}

Compared to run \runa, which has otherwise the same parameters, the profiles
of the kinetic and magnetic energies, seen in Fig.~\ref{fig:resp}, are
barely affected: such small dissipation coefficients, which allow for a
wide turbulent inertial range to develop, have almost no influence on the
energy content as a function of the position along the loop, and the energy
does not decrease significantly with altitude as a result of the
dissipation.

However, as expected, the profiles of both the kinematic viscosity $\nu$ and
the magnetic diffusivity $\eta$ have an effect on the kinetic ($H_V^u$) and
magnetic ($H_V^b$) dissipation powers (right panels of Fig.~\ref{fig:resp}):
the profile of $H_V^u$ is mainly affected by the profile of $\nu$ and the
profile of $H_V^b$ is mainly affected by the profile of $\eta$.  The
combination of both gives a heating profile which is enhanced either at the
footpoints (runs \runp\ and \runq), at the loop top (runs \runs\ and \runt)
or both (run \runr).  Heating enhancements at the footpoints are mainly due
to magnetic energy dissipation, while a heating enhancement at the loop top
is mainly due to kinetic energy dissipation.  Note also that, as the kinetic
energy is smaller in all cases than the magnetic energy, the ratio of the
kinetic over magnetic dissipation is lower than what would be expected from
the mere ratio of the kinematic viscosity over the magnetic diffusivity.

The fits of the kinetic (resp. magnetic) dissipation profiles to a power-law
$\rho^\alpha$ are in most cases broadly consistent with the expected value
$\alpha = -7/2$ (resp. $3/2$), although there are important differences
between the different runs.  The ratios between the kinetic (resp. magnetic)
dissipation power near the footpoints and at the loop top are also broadly
consistent with the same ratio for $\nu$ (resp. $\eta$).  Given the
parameters of the model, this ratio can become very large but it is again
easily managed by the \shellatm\ model.

\section{Discussion}
\label{sec:disc}

We have presented a set of simulations of a coronal magnetic loop containing
a highly turbulent MHD plasma, with Reynolds numbers up to $10^6$, a
variation between the loop top and the footpoints by a factor of up to $30$
for the density and up to the order of $10^7$ for the magnetic Prandtl
number.  These ranges of physical parameters cannot be reached by direct
numerical simulations.

The spatial distributions of the energy and of the heating as a function of
the position along the loop have been obtained.  The profiles of energy do
not have a strong dependence on the position along the loop (runs \runa\ to
\runc), except when the flux tube is expanding (runs \runA\ to \runC): in
this case the energy per unit volume is higher near the footpoints, where
the magnetic field is higher.  The profiles of the heating (the dissipation
of energy per unit volume) also have a low dependence on position when only
the density varies.  If the flux tube is expanding, the behavior of the
heating profile is quite similar to that of the energy profile: the heating
is higher near the footpoints.

When we look at runs \runa\ to \runC\ the profiles of energy and heating are
shallower than what is expected from the simple linear phenomenology
introduced in Sec.~\ref{sec:phen}.  We now try to explain the origin of
these discrepancies.  First, one might expect that the dissipation occurring
during the wave propagation to the loop top would make the profile of energy
$E_V$ deeper instead of shallower; however, the comparison of the energy
profiles in runs \runp\ to \runt\ (Fig.~\ref{fig:resp} left) with run \runa\
shows that this effect remains unnoticeable.  The origin of the energy
profile is still unclear.

Assuming now an energy profile $E_V \propto \rho^{\alpha_E}$ (with
$\alpha_E$ coming from the fits of the results of the different runs), the
dissipation profile in a given shell $n$ is $H_V^{(n)} \propto B_\parallel
\rho^{\alpha_E}$, i.e. $\alpha_H \approx 0.1$ for runs \runa\ to \runc\ and
$\alpha_H \approx 0.55$ for runs \runA\ to \runC\footnote{It can
    also be noted that these new exponents are close to the exponents
    $\alpha_H=0$ and $\alpha_H=1/2$ that would be obtained from a heating
    proportional to $B_{\parallel}^2$ \citep{gudiksen05b}.}.  This is
already closer to the simulations results (Table~\ref{tab:expind}) than the
purely phenomenological predictions of $1/2$ and $3/4$ respectively.  The
remainder of the difference could be explained by the effect observed in run
\runu: the ratio of the kinetic dissipation profile (which is low at
footpoints due to the velocity boundary condition at the photosphere) over
the magnetic dissipation profile is increased compared to the ratio of the
kinetic energy profile over the magnetic energy profile because of the
concentration of magnetic energy at the larger scales in the loop.  Thus the
slow motions of the loops footpoints anchored in the photosphere and the
build-up of magnetic energy mainly at large scales have a decisive influence
on the determination of the profiles of heating in coronal loops.

Another, more subtle, effect is related to the profiles of the nonlinear and
dissipation timescales.  For a given $\kperp$, the non-linear time scale is
$\tnl = 1 / \kperp Z$ and the dissipation time scale is $\tnu = 1 / \nu
\kperp^2$.  Taking into account the energy profile and the variation of
$\kperp$ for a given shell $n$ (like in Sec.~\ref{sec:phen}), the non-linear
time scale is $\tnl^{(n)} \propto \rho^{(1-\alpha_E)/2} / B_\parallel^{1/2}$
and the dissipation time scale is $\tnu^{(n)} \propto 1 / \nu B_\parallel$
for a given shell $n$.  This means that for runs \runa\ to \runc, the
dissipation time scale does not depend on position while the nonlinear
timescale ($\approx \rho^{0.45}$) is shorter at the loop top: the transfer
of energy to the small scales is more efficient, and as a consequence, the
dissipation power is higher at the loop top than what is expected from the
phenomenology of Sec.~\ref{sec:phen}.  This is in accordance to what is seen
in the simulation results (a shallower profile of dissipation).  For runs
\runA\ to \runC\ the nonlinear ($\propto \rho^{0.23}$) and dissipation
($\propto \rho^{-1/4}$) timescales vary in opposite directions: the
nonlinear terms are more efficient at the loop top, but they need to bring
the energy further in the spectrum.  However, the different
$k_\perp$-dependence of both competing timescales means that the enhanced
efficiency of the nonlinear transfer ``wins'' over the displacement of the
dissipation scale: again, as seen in the simulations, the heating profile is
shallower than expected before.

Runs \runp\ to \runt\ point out the additional role of the variation of the
diffusion coefficients as a function of temperature (and thus as
a function of position along the loop), which seems to have been overlooked
in previous works.  The profile of the kinetic (magnetic) energy dissipation
follows approximately the profile of the kinetic (magnetic) diffusion
coefficient respectively.  The diffusion coefficients have thus a direct
influence on the energy dissipation, and as they strongly vary with
temperature, this yields a strong variation of the dissipation power along
the loop.  Furthermore, as both diffusion coefficients vary in opposite ways
with temperature, the heating can be enhanced at the footpoints, at the loop
top or both, depending on the magnetic Prandtl numbers in the loop.

Although these results seem to be straight forward, this had never been
modelled in the context of solar coronal heating, as the precise values of
$\nu$ and $\eta$ are usually considered unimportant to get averages of the
heating: the main argument for that is that energy is dissipated at the end
of the inertial range, at whichever scale (or wavenumber) this end is
\citep{gals971}.  However, these values actually have an importance because
(1) they discriminate between low- and high-Reynolds-number physics, with
effects like intermittency appearing only at very high Reynolds numbers; (2)
when waves propagate between regions with different diffusion coefficients,
we have shown in this paper that the profiles of these coefficients
contribute to the determination of the profile of the heating.

These simple simulations shed light on the consequences on heating profiles
of the complex interactions between some of the physical effects that come
into play in a non-uniform turbulent coronal loop: conservation of energy
and magnetic fluxes, accumulation of large-scale magnetic field in a loop
submitted to photospheric motions, nonlinear effects, and non-uniformity of
the viscosity and magnetic diffusivity coefficients.  However, many
processes have not been included in these simulations, such as heat
transport (conduction and radiation), flows, gravity, MHD processes
overlooked by the shell-models, and kinetic effects.  Future developments
may include the heat transport and the subsequent computation of emitted
radiation, especially in UV.  In addition to completing the picture of the
nanoflare-like processes involved in heating and cooling of loops, this will
allow the comparison of such models to data from new instruments (such as
Hinode/EIS) that observe the plasma at higher temperatures than previous
instruments, at earlier stages of heating events.

\begin{acknowledgements}
  Collaboration with Andrea Verdini when developing the numerical code is
  greatly acknowledged.  EB acknowledges support from a PPARC rolling grant.
  SJB is grateful to PPARC for the award of a Post-Doctoral Fellowship.  MV
  acknowledges support from NASA grant SHP04-0000-0150.  The authors thank
  the referee, D.A.N. Müller, for intersting comments that helped improve
  the manuscript.
\end{acknowledgements}

\end{document}